\documentclass[aps,amssymb,prb,twocolumn,showpacs]{revtex4}
\usepackage{graphicx}
\usepackage{epsfig,amsmath}
\pdfoutput=1
\begin{document}

\title{Electromechanical instability in vibrating quantum dots with effectively negative charging energy}

\author{Teemu Ojanen}
\email[Correspondence to ]{ojanen@physik.fu-berlin.de}
\author{Friedrich C. Gethmann}
\author{Felix von Oppen}
\affiliation{Institut f\"ur Theoretische Physik, Freie Universit\"at Berlin,
Arnimallee 14, 14195 Berlin, Germany}
\date{\today}
\begin{abstract}
In quantum dots or molecules with vibrational degrees of freedom the electron-vibron coupling renormalizes the electronic charging energy. For sufficiently strong coupling, the renormalized charging energy can become negative.
Here, we discuss an instability towards adding or removing an arbitrary number of electrons when the magnitude of the renormalized charging energy exceeds the single-particle level spacing. We show that the instability is regularized by the anharmonic contribution to the vibron energy. The resulting effective charging energy as a function of the electron number has a double-well structure causing a variety of novel features in the Coulomb blockade properties.

\end{abstract}
 \bigskip

\maketitle

\section{Introduction}

The interplay of mechanical motion and electronic properties has been studied widely in molecular physics and, recently, in molecular electronics and nanomechanical structures. Present-day experimental methods enable the fabrication of nanostructures such as single-molecule junctions or quantum dots where a small number of electrons couple to one or more vibrational modes. Electronic transport in these systems exhibits vibrational sidebands at high bias voltages,\cite{mitra,braig,mozyrsky,weig} the shuttle instability,\cite{gorelik,novotny,pistolesi2,jonsson} and the Franck-Condon suppression of the low-bias electric conductance.\cite{koch2,koch3,leturcq} The shuttle phenomenon arises from modifications in the spatial conformation driven by the charging events, whereas the Franck-Condon effects emerge from the transition rates between discrete vibronic states. An additional important consequence of vibrational degrees of freedom is the polaron shift which renormalizes the Coulomb repulsion of confined electrons and modifies the Coulomb blockade physics of the system. At sufficiently strong couplings the sign of the effective charging energy $U$ can become negative implying an effectively attractive interaction between electrons.\cite{alexandrov,cornaglia,arrachea,mravlje,koch1,koch4} The regime of a negative effective charging energy is known to be realized in certain molecules (an effect known in chemistry as potential inversion).\cite{kraiya} It may also be accessible in vibrating quantum dots when the electron-vibron coupling is sufficiently strong.\footnote{Strong electron-vibron coupling has been observed recently, e.g., in suspended carbon nanotube quantum dots. \cite{leroy,sapmaz,kuemmeth,huttel, leturcq}}

It has been shown recently that the negative $U$ regime opens the possibility for novel features such as an efficient electron pair tunneling through single molecules \cite{koch1} and a charge-Kondo effect.\cite{koch4} An underlying assumption of these works is that the negative charging energy does not induce an instability of the system. This is indeed the case when the energy gain due to the effectively attractive charging energy is smaller than the cost in single-particle energy due to the finite level spacing, when adding (removing) electrons to (from) the dot. This paper is devoted to study the opposite regime where the magnitude of the negative (renormalized) charging energy exceeds the single-particle level spacing. The studied system is modeled as a single-electron transistor (SET) whose center island is coupled to a mechanical vibration.\cite{park,sazonova,koch2, sapmaz,naik,pistolesi1} It is shown that the system becomes unstable towards addition or extraction of electrons. Assuming that the vibron Hamiltonian contains an anharmonic contribution we show that the instability is regularized and that the system possesses a well defined ground state. The effective charging energy as a function of the electron number in the dot has a double-well structure which, moreover, depends on temperature. This is in striking contrast to the usual parabolic charging energy and leads to a variety of novel effects. At zero gate charge and low temperatures, there is a symmetry between particle and hole-like excitations causing the existence of degenerate minima of the effective charging energy which for weak anharmonicity are separated by a large number of electrons. The average population and the electron number fluctuations are highly sensitive to the gate voltage which breaks the symmetry between the minima. We also show that the transport properties of the system exhibit a number of distinctive features attributed to the unusual form of the effective charging energy.

This paper is organized as follows. In Sec.\ \ref{model} we introduce the model, demonstrate the origin of the negative-$U$ instability, and state the precise conditions under which it takes place. In Sec.\ \ref{anharmonic} we explore the equilibrium properties of the instability, including the effects of the vibron nonlinearity. Nonequilibrium (transport) properties in the presence of a bias voltage are discussed in Sec.\ \ref{transport}. We conclude in Sec.\ \ref{conclusions}.

\section{Model and negative $U$ instability} \label{model}

\begin{figure}[h]
\begin{center}
\includegraphics[width=0.7\columnwidth]{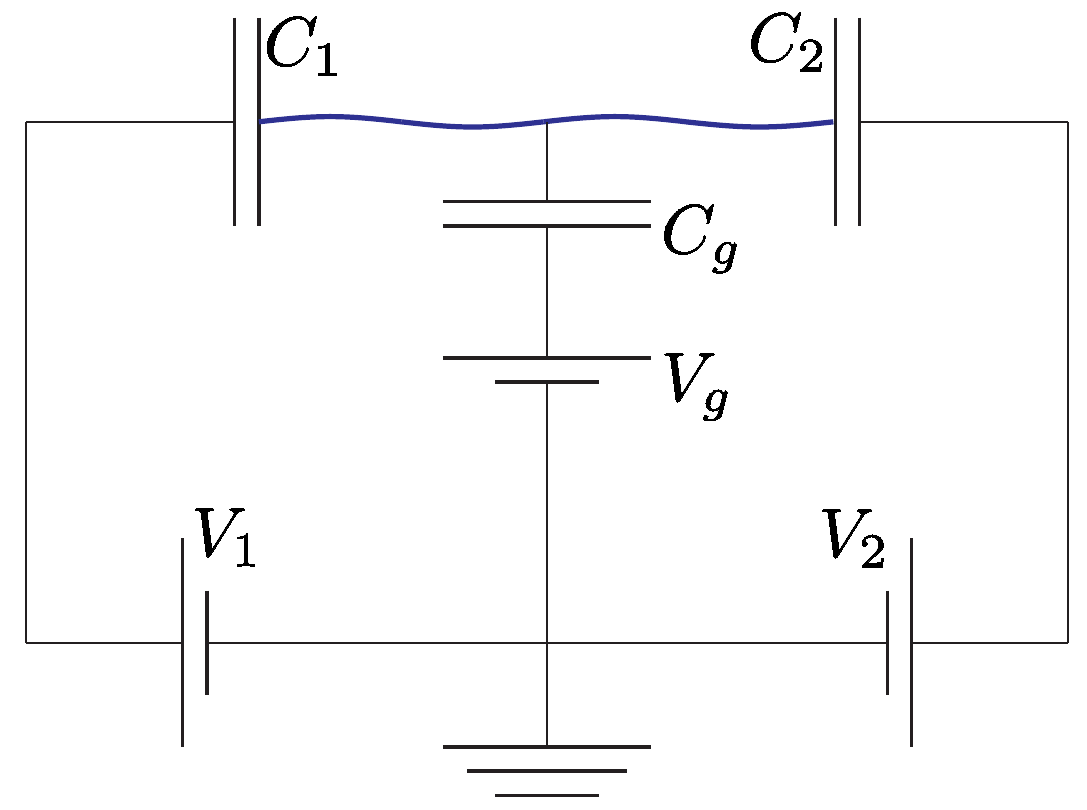}
\caption{Equivalent circuit of the quantum dot. Capacitances $C_1$ and $C_2$ model high resistance tunnel junctions which allow the island
to exchange electrons with the reservoirs, $C_g$ represents a gate capacitance. The central part is coupled to a vibrational mode.}
\label{Abb_1}
\end{center}
\end{figure}
We are considering a quantum dot or a single-molecule junction coupled to a single vibrational degree of freedom. The dot is modeled as a SET where the number of electrons on the island couples to a vibron mode, see Fig.~\ref{Abb_1}. The system could be realized, for example, by a suspended center element between large metallic leads enabling tunneling between the dot and the leads. The electronic part of the Hamiltonian for a SET is
\begin{align}\label{eham}
H_{\rm e}=\sum_{\rm L,R,C}\epsilon_jc_j^{\dagger}c_j+H_{\rm T}+U(N),
\end{align}
where the first term represents the single-particle energies of the leads and the center part, the second term describes tunneling between the leads and the center, and the last term is the relevant charging energy of the system. The tunneling Hamiltonian and the charging energy take the standard forms
 \begin{align}
H_{\rm T}=\sum_{i\in {\rm C}, j\in {\rm L/R}} \left(t_{ij}c_i^{\dagger}c_j+t_{ij}^*c_j^{\dagger}c_i\right),
\end{align}
and
\begin{align}
U(N)=\frac{(Ne-Q_g)^2}{2C},
\end{align}
respectively. Here, $N=\sum_{\rm C}c_j^{\dagger}c_j-N_0$ is the number of extra electrons on the island, $C=C_1+C_2+C_g$, and $Q_g=V_1C_1+V_2C_2+V_gC_g$ is the total induced gate charge. The number $N_0$ corresponds to the positively charged lattice ions on the central island. The full Hamiltonian of the system $H=H_{\rm e}+H_{\rm ph}$ also includes the vibrational contribution
\begin{align}\label{pham}
\!\!\!\!H_{\rm ph}=\frac{P^2}{2M}+\frac{1}{2}M\Omega^2x^2+\alpha(2M\Omega^2)^2x^4\!+\!\lambda\, \hbar\Omega\, N\frac{x}{l_{\rm osc}}
\end{align}
where we have included the electron-vibron coupling $\lambda\, \hbar\Omega\, Nx/l_{\rm osc}$ with $l_{\rm osc}=\sqrt{\hbar/2M\Omega}$ into the phonon Hamiltonian. The Hamiltonian (\ref{pham}) also includes an anharmonic $x^4$ term whose strength is controlled by the parameter $\alpha$. The anharmonic term becomes comparable to harmonic effects when the phonon displacement is of the order of $(\alpha\hbar\Omega)^{-2}$ times the oscillator length $l_{\rm osc}$. The inclusion of this term, even in the weakly nonlinear case $\alpha\,\hbar\Omega\ll1$, is crucial for the stability of the system for large couplings $\lambda$. To demonstrate this let us first assume $\alpha=0$ and examine the phonon-induced modification of the charging energy. Since the electron number couples linearly to the phonon position, it is possible to introduce a shifted phonon coordinate by completing the square in the phonon Hamiltonian. As a result one eliminates the coupling term and generates an extra term $\Delta U(N)$ quadratic in $N$ which can be thought of as a modification of the charging energy
\begin{align}\label{epot}
 U^{\rm eff}(N)&=U(N)+\Delta U(N)=\frac{(Ne-Q_g)^2}{2C}-N^2\lambda^2\hbar\Omega.
\end{align}
It is important to recognize that the well-known polaron shift $\Delta U(N)$ gives rise to a \emph {negative} contribution.
The overall prefactor of the $N^2$-term will be negative if
\begin{align} \label{neg}
\gamma=\frac{e^2}{2C}-\lambda^2\hbar\Omega<0,
\end{align}
meaning that the effective potential favors large absolute values of $N$. Defining the quantity $E_C=e^2/2C$, the negative charging energy condition (\ref{neg}) takes the form
\begin{align}\label{cond1}
\lambda>\sqrt{\frac{E_C}{\hbar\Omega}}\,.
\end{align}
Consequently, in metallic quantum dots with a vanishingly small single-particle level separation, the  energy can always be lowered by adding or removing an arbitrary number of particles when condition (\ref{cond1}) holds. Thus the system does not possess a ground state and is unstable. In dots whose orbital energies due to confinement are not negligible compared to the charging and vibron energies, the single-particle energy cost should also be taken into account. To analyze this, we introduce equidistant single-electron orbitals
\begin{equation}\label{singorb}
E_n=\hbar\,v_{\rm F}\frac{n\pi}{L_d}=\Delta n,
\end{equation}
where $v_{\rm F}$ is the Fermi velocity of the dot, $L_d$ is the effective size of the electronic confinement and $\Delta=\hbar v_ {\rm F}\pi/L_d$ is the level separation. Expression (\ref{singorb}) is appropriate in the vicinity of the Fermi energy and holds exactly for a one-dimensional Dirac spectrum. The energy cost of adding or removing $N$ electrons to or from the lowest available orbitals (ignoring the spin degeneracy) is
\begin{equation}\label{singcost}
\sum_{n=1}^N E_n=\Delta\frac{N(N+1)}{2}.
\end{equation}
Adding the single-particle contribution to the effective potential (\ref{epot}) we get
\begin{align}\label{epotmod1}
 &U^{\rm eff}(N)=\Delta\frac{N(N+1)}{2}+E_C(N-Q_g/e)^2-N^2\lambda^2\hbar\Omega\nonumber\\
 &=\left(\frac{\Delta}{2}+E_C\right)(N-Q_g/e)^2-N^2\lambda^2\hbar\Omega+{\rm const},
\end{align}
where the second line follows by shifting the gate charge $Q_g\to Q_g+\frac{\Delta}{2E_C}(\frac{e}{2}+Q_g)$. The constant in Eq.~(\ref{epotmod1}) is independent of $N$ and can be dropped. Thus the functional form of the effective potential remains invariant, and the finite level separation just renormalizes charging energy and gate charge. The system is still fundamentally unstable provided that
\begin{align}\label{cond2}
\lambda>\sqrt{\frac{E_C+\Delta/2}{\hbar\Omega}}.
\end{align}
In the spin-degenerate case $\Delta$ should be replaced by $\Delta/2$. Alternatively, we can write this condition as
$E_C-\lambda^2 \hbar\Omega < -\Delta/2$, i.e., we find an instability when the magnitude of the renormalized charging energy becomes larger than the level spacing.

Result (\ref{cond2}) shows that a constant (or decreasing) level spacing is not adequate to stabilize the system at sufficiently large couplings. The instability tends to change the electron number by increasing the displacement of the phonon mode. Therefore it is natural to assume that at some point when the displacement becomes large the harmonic approximation for the vibron mode breaks down and anharmonic effects become significant. It turns out that a generalization of the above analysis to include a nonzero anharmonic term ($\alpha>0$) in Eq.~(\ref{pham}) always produces a well-defined ground state. The effective potential for electrons becomes temperature dependent and exhibits a variety of novel features.

Our discussion has a close relation to familiar stability conditions in Fermi-liquid theory.\cite{pines} In fact, it is well known that the Fermi liquid remains stable even when the Landau interaction parameter $F_0$ becomes negative as long as
$F_0>-1$. Since $F_0$ measures the interaction strength in units of the density of states, this is a precise analog of the stability condition (\ref{cond2}) expressed as $(E_C-\lambda^2\hbar\Omega)/(\Delta/2)>-1$.\footnote{Quantity $(E_C-\lambda^2\hbar\Omega)/(\Delta/2)$ corresponds to $F_0$ apart the difference of factor two which can be readily traced to trivial differences in definitions.}

\section{Thermal equilibrium properties} \label{negative}
\label{anharmonic}

\subsection{Effective potential}

In this section we study effects of the anharmonic phonon term on the negative-$U$ instability, focusing on thermal equilibrium properties of the junction. Diagonalizing (\ref{pham}) one obtains phonon eigenvalues $\{E_j(N)\}$ as a function of the electron number $N$ and the island partition function becomes \footnote{For simplicity, we present our analysis in the case $\Delta=0$. A nonzero $\Delta$ can be readily included by a simple renormalization of $E_C$ and $Q_g$. }
\begin{eqnarray}
Z&=&\sum_{N,j}\mathrm{exp}\left\{-\beta\left[U(N)+E_j(N)\right]\right\} \nonumber \\
&=&\sum_{N}\mathrm{exp}\left\{-\beta U(N) \right\}
\sum_{j}\mathrm{exp}\left\{-\beta E_j(N)\right\}.
\label{part}
\end{eqnarray}
In Eq.~(\ref{part}) we have assumed that the electron tunneling is weak and its contribution to the partition function is negligible compared to the charging energy. \cite{note}
Defining a phonon partition function as $Z_{\rm ph}(N)=\sum_{j}\mathrm{exp}\left\{-\beta E_j(N)\right\}$, the partition function takes the form
\begin{align}\label{part1}
& Z=Z_{\rm ph}(N=0)\nonumber\\
&\times\sum_{N}\mathrm{exp}\left\{-\beta[U(N)-\frac{1}{\beta}\mathrm{ln}\left(Z_{\rm ph}(N)/Z_{\rm ph}(N=0)\right)]\right\}.
\end{align}
From expression $(\ref{part1})$ one can identify the effective electronic potential
\begin{equation}\label{effpot}
U^{\mathrm{eff}}(N,T)= U(N)-\frac{1}{\beta}\mathrm{ln}\,\left[Z_{\rm ph}(N)/Z_{\rm ph}(N=0)\right].
\end{equation}
In the absence of the electron-phonon interaction, Eq.\ (\ref{effpot}) reduces to the ordinary charging energy, and in the absence of the anharmonic term $(\alpha=0)$ it is temperature independent and coincides with expression (\ref{epot}). In the anharmonic case the polaron shift depends on the phonon state and thus the effective potential depends on the phonon temperature. In general, an evaluation of Eq.\ (\ref{effpot}) requires one to diagonalize (\ref{pham}) numerically and to compute the phonon partition function. The behavior of the system is then determined by the relative strength of the energy scales $E_C$, $\hbar\,\Omega$, $\alpha^{-1}$, and the magnitude of dimensionless electron-phonon coupling $\lambda$.

\subsection{Analytical considerations}
\label{analytical}

Analytical expressions for the effective potential can be obtained for weakly unstable systems at zero temperature when considering the vibronic degree of freedom as classical. As this already illustrates some of the essential physics, we first consider this case before presenting more general numerical results for the fully quantum mechanical model.

Starting with Eq.\ (\ref{effpot}), we can write the effective potential at zero temperature as
\begin{equation}
  U^{\rm eff}(N) = U(N) + E_0(N) - E_0(0),
\label{effpotzero}
\end{equation}
where $E_0(N)$ denotes the ground state energy of the phonon Hamiltonian $H_{\rm ph}$ for excess charge $N$. In the classical limit, the ground state energy is simply given by the minimum of the potential energy. Noting that $E_0(0)$ vanishes, we obtain
\begin{align}
  U^{\rm eff}&(N) = U(N) - \lambda^2 \hbar \Omega N^2 \nonumber\\
  &+ \min_x \{\frac{1}{2}M\Omega^2 (x+2\lambda N \ell_{\rm osc})^2
   +\alpha(2M\Omega^2)^2x^4 \}.
   \label{effpotmin}
\end{align}
For a weakly unstable system near zero gate charge, we expect that the effective potential has a minimum for small (but nonzero) $N$. Thus, we can neglect the quartic term when determining the position of the minimum in Eq.\ (\ref{effpotmin}),
\begin{equation}
  x_0 \simeq -2\hbar\Omega N.
\end{equation}
It is straightforward to include corrections to this expression and find that this approximation is valid as long as $\lambda^2 |E_C-\lambda^2\hbar\Omega|/\hbar \Omega\ll 1$, i.e., as long as the instability is only weakly developed. With this expression for $x_0$, we find for the effective potential
\begin{equation}
  U^{\rm eff}(N) = U(N) - \lambda^2 \hbar \Omega N^2 + 16 \alpha (\hbar\Omega)^2 N^4.
\label{effpotcl}
\end{equation}
Focusing on the case of zero gate charge, we can now compute the number of electrons $N_0$ which are entering the quantum dot as a consequence of the instability. Minimizing the effective potential in Eq.\ (\ref{effpotcl}), we find
\begin{equation}
  N_0^2 = \frac{|E_C - \lambda^2 \hbar \Omega |}{32 \alpha (\hbar\Omega)^2}.
\end{equation}
Note that there are indeed degenerate minima at $N=\pm N_0$ for zero gate charge.

Once the instability develops, we can define an effective charging energy $E_C^{\rm eff}$ which decribes the curvature of the effective potential in the vicinity of the minima. Expanding Eq.\ (\ref{effpotcl}) around the minima, we find
\begin{equation}\label{curvature}
    E_C^{\rm eff} = 2 |E_C-\lambda^2 \hbar \Omega|
\end{equation}
This sign reversal of the effective charging energy induced by the instability has important consequences. While a negative renormalized charging energy leads to pair tunneling in the absence of the instability,\cite{koch2} the positive effective charging energy $E_C^{\rm eff}$ implies that transport will be dominated again by single electron sequential tunneling processes once the instability develops. It is also interesting to note that the Coulomb blockade is significantly weakened near the instability where many charge states have similar energies.

We close this section with a remark on the validity of the classical approximation for the vibrons employed in this section. As can be readily seen from Eq.\ (\ref{effpotzero}), the quantum-mechanical zero-point energy of the vibron mode cancels out from this expression as long as the vibron remains harmonic, implying that the zero-point energy is independent of the electron number $N$. Thus, quantum corrections to our classical discussion are proportional to the anharmonicity and our classical expressions are an excellent approximation to the results of a fully quantum mechanical calculation as long as the anharmonicity remains small (see Sec.\ \ref{numerical} for a comparison with numerical results).

\begin{figure}[h]
\begin{center}
\includegraphics[width=0.99\columnwidth]{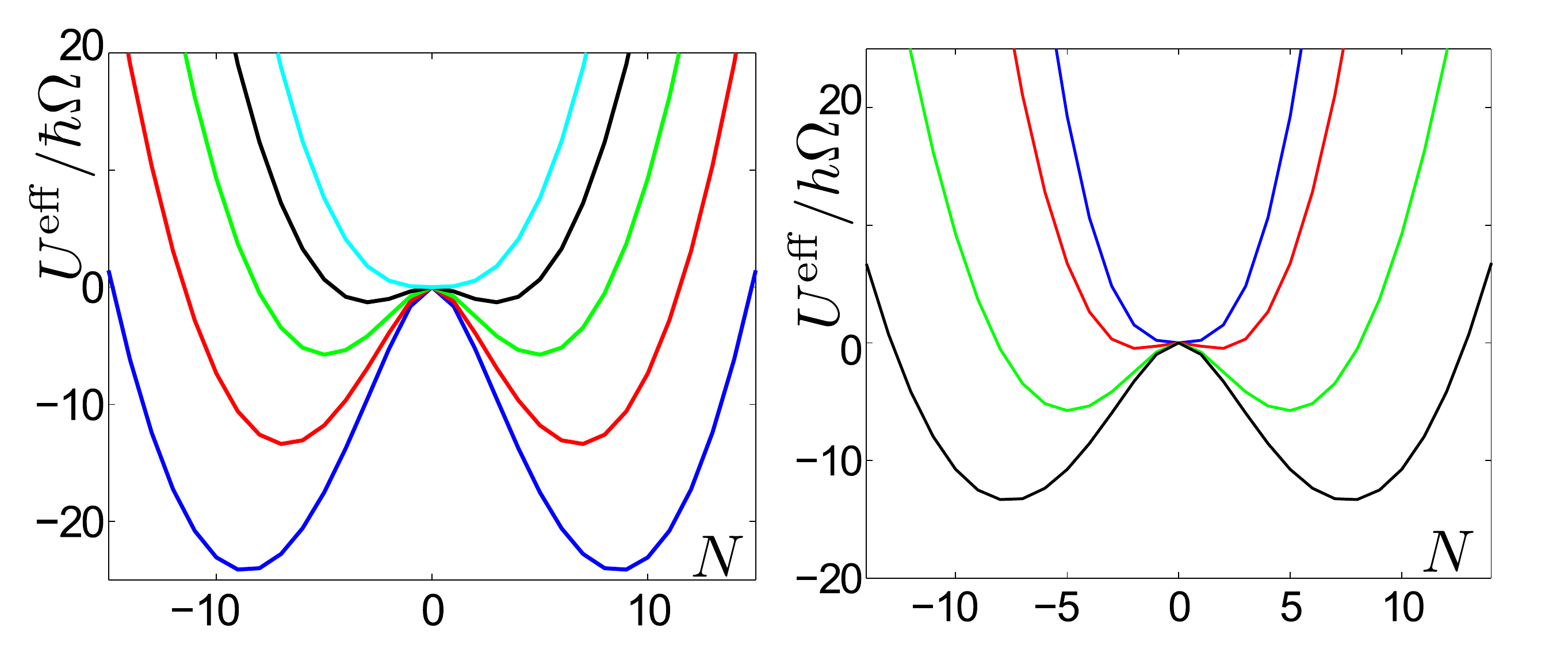}
\caption{Effective potential as a function of the electron number. Left: the coupling-dependence of the potential, $T=0.1\,\hbar\Omega$, $\alpha=0.01$, $E_C=\hbar\Omega$, $Q_g=0$ and from bottom to top $\lambda^2=3,\, 2.5,\, 2,\, 1.5,\,1$. Right: the effective potential corresponding to different charging energies, $\alpha=0.01(\hbar\Omega)^{-1}$, $\lambda^2=2$, $T=0.1\hbar\Omega$, $Q_g=0$, and from bottom to top $E_C=0.5,\,0.8,\,1,\,1.5,\,2\times \hbar\Omega$. }
\label{morepot}
\end{center}
\end{figure}

\subsection{Numerical results}
\label{numerical}

A solution of the full quantum problem for arbitrary parameter values requires a numerical approach. We now focus on the results of such a calculation. Figure \ref{morepot} illustrates the dependence of the effective potential on the electron-phonon coupling strength and the charging energy. As the coupling strength is increased the system exhibits a crossover between the $\gamma>0$ regime and the negative-$U$ regime $\gamma<0$, signalled by the formation of a double-well structure. Decreasing the (bare Coulomb) charging energy results in qualitatively similar behavior as increasing the coupling strength.
\begin{figure}[h]
\begin{center}
\includegraphics[width=0.99\columnwidth]{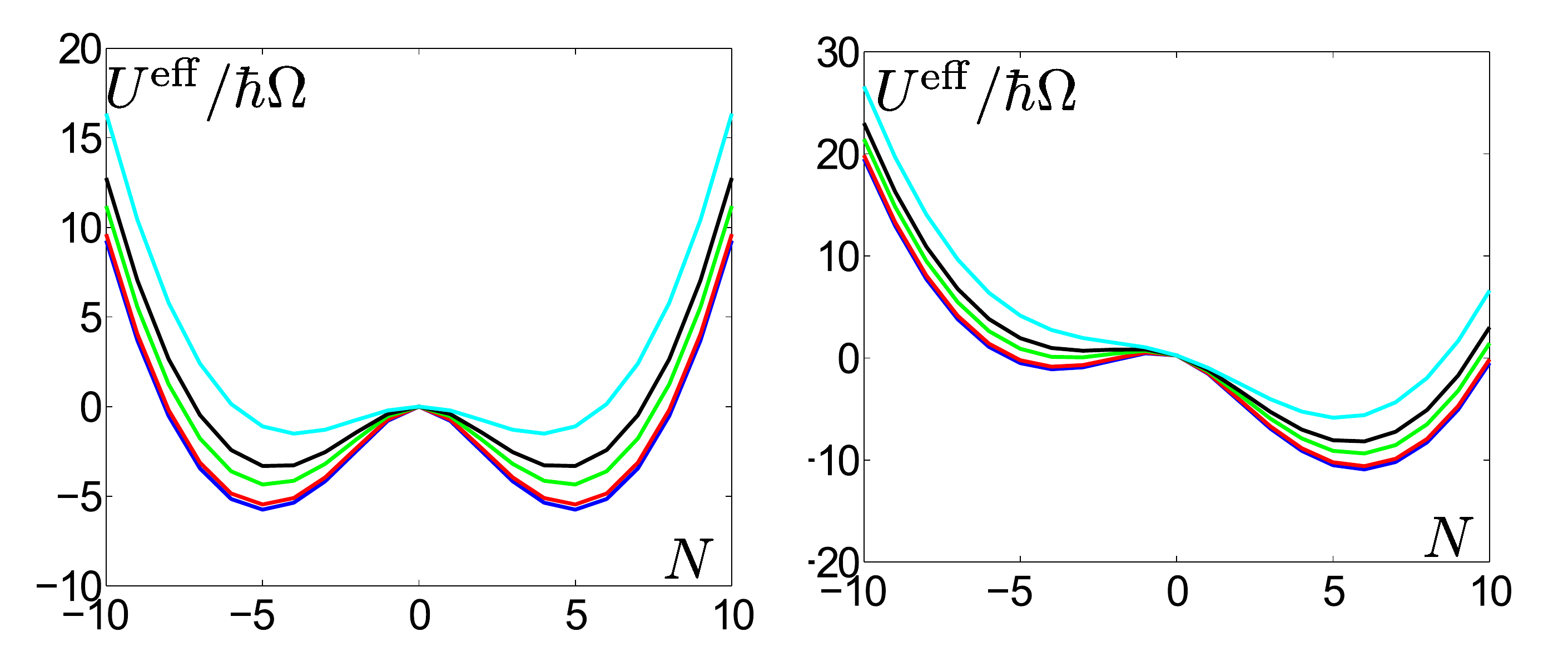}
\caption{Effective potential as a function of the electron number. Left: Curves illustrate the temperature dependence of the potential corresponding to parameters $E_C=\hbar\Omega$, $\alpha=0.01(\hbar\Omega)^{-1}$, $\lambda^2=2$, $Q_g=0$ at $T=0.1,\,3,\,5,\, 10\times\hbar\Omega$ (from bottom to top ). Right: The same as the left figure but at finite gate charge $Q_g=e/2$.}
\label{efftemp}
\end{center}
\end{figure}

\begin{figure}[h]
\begin{center}
\includegraphics[width=0.99\columnwidth]{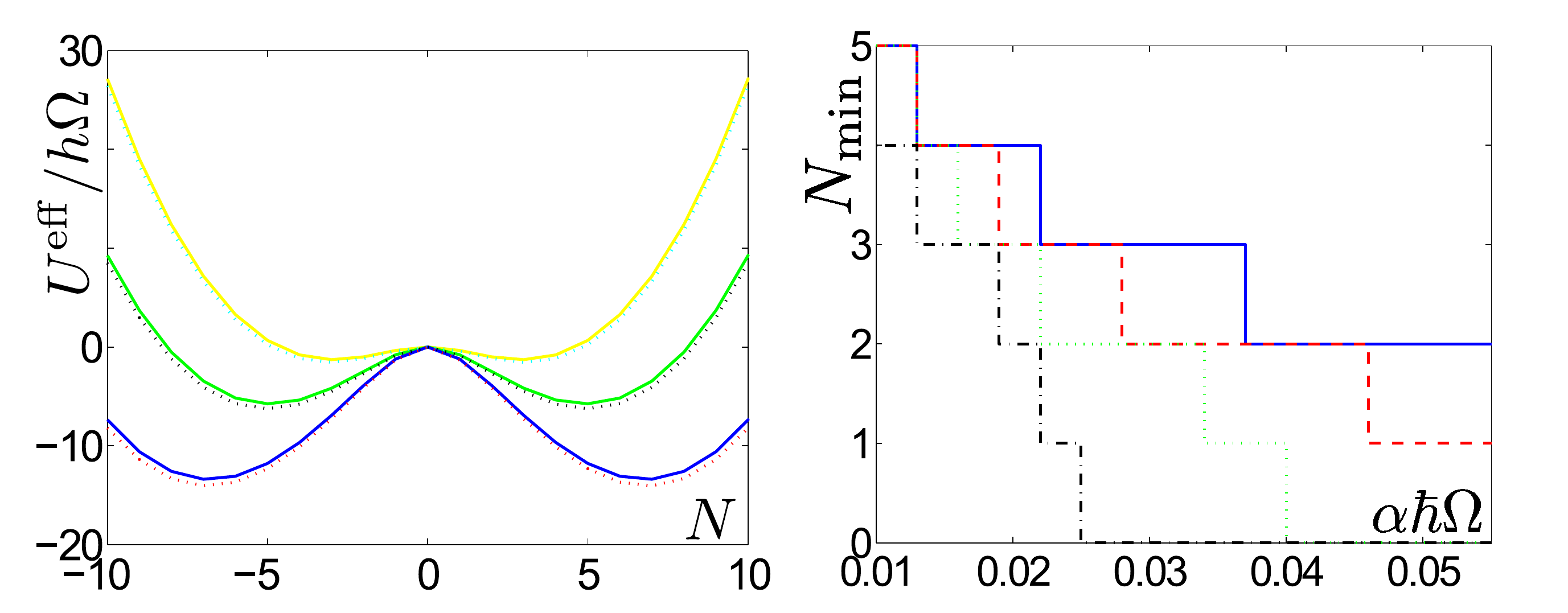}
\caption{Left: Comparison of the classical and quantum mechanical effective potentials. Solid curves correspond to the quantum mechanical calculation with parameters $E_C=\hbar\Omega$, $\alpha=0.01(\hbar\Omega)^{-1}$, $Q_g=0$, $T=0$ and $\lambda^2=2.5$ (blue), $\lambda^2=2.0$ (green), $\lambda^2=1.5$ (yellow). The dotted lines correspond to the classical approximation obtained by minimizing the phonon potential energy. Right: Electron number corresponding to the minimum of the right potential well as a function of the nonlinearity,  $E_C=\hbar\Omega$, $\lambda^2=2$, $Q_g=0$ and $T=0.1\hbar\Omega$ (solid), $T=3\hbar\Omega$ (dashed), $T=5\hbar\Omega$ (dotted) and $T=10\hbar\Omega$ (dash-dotted).}
\label{q_vs_cl}
\end{center}
\end{figure}
Figure \ref{efftemp} shows the temperature dependence of a typical effective charging curve $U^{\mathrm{eff}}(N,T)$ in the negative-$U$ regime $\gamma<0$. At low temperatures the potential exhibits a double-well structure whose details depend on the particular parameter values. The shape of the potential can also change qualitatively as the temperature is increased: the double well structure present in Fig.~\ref{efftemp} is eventually deformed to a single-well potential. The temperature dependence of the potential is a consequence of the nonlinear phonon interaction. In the asymmetric case (i.e., at nonzero gate charge $Q_g\neq 0$) there exists a metastable energy minimum whose decay towards the global minimum requires multiple electron tunneling (roughly 10 electrons for the parameter values). Figure \ref{q_vs_cl} (left) shows a comparison between the full quantum mechanical solution obtained numerically and the classical result obtained in Se.\ \ref{analytical}. As argued above, we find that the classical approximation is quite accurate for small anharmonicity. The wells are closer at stronger anharmonicity and higher temperature, as indicated by Fig.~\ref{q_vs_cl} (right).

\begin{figure}[h]
\begin{center}
\includegraphics[width=0.99\columnwidth]{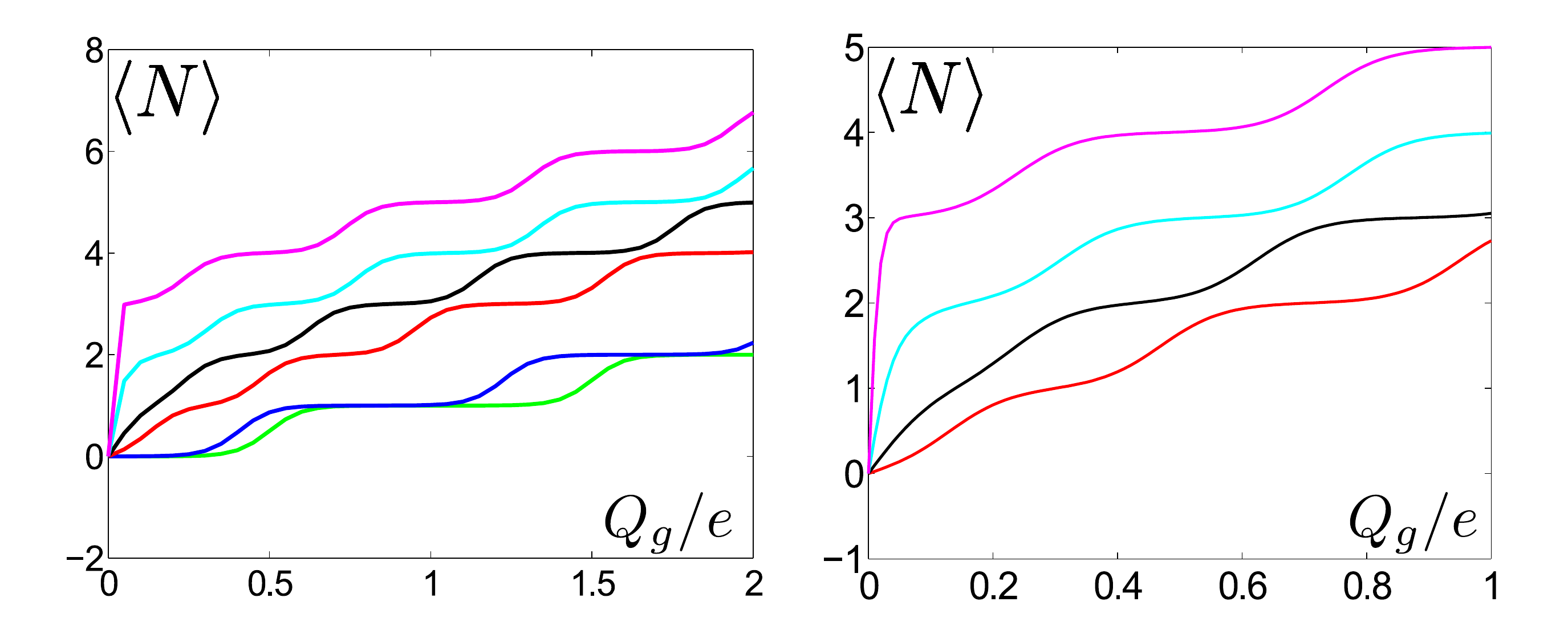}
\caption{Average dot population as a function of the gate charge. The left figure corresponds to parameters $T=0.1\,\hbar\Omega$, $\alpha=0.01(\hbar\Omega)^{-1}$, $E_C=\hbar\Omega$ and from bottom to top $\lambda^2=0,0.2,0.8, 1, 1.2$ and $1.5$. At vanishing coupling ($\lambda=0$) we recover the usual Coulomb steps, in the strong coupling we have a rapid increase at low gate voltages. The right figure clarifies the small gate charge region for couplings $\lambda^2=0.8,\, 1,\, 1.2,\,1.5$ (from bottom to top).}
\label{ngates}
\end{center}
\end{figure}
The average number of electrons as function of the gate charge is presented in Fig.~\ref{ngates}. In the absence of the electron-phonon coupling and at low temperatures, the average number $\langle N\rangle$ exhibits the usual Coulomb staircase behavior. As the coupling is increased, this is gradually transformed into a new dependence reflecting the double-well nature of the effective potential. The double-well structure leads to a rapid increase of the population at low gate charges. The reason for this is that the gate acts as a symmetry breaking field and the values $Q_g/e\gtrsim k_{\rm B}T/E_C$ are sufficient for the system to prefer one of the two nearly degenerate wells.

\begin{figure}[h]
\begin{center}
\includegraphics[width=0.99\columnwidth]{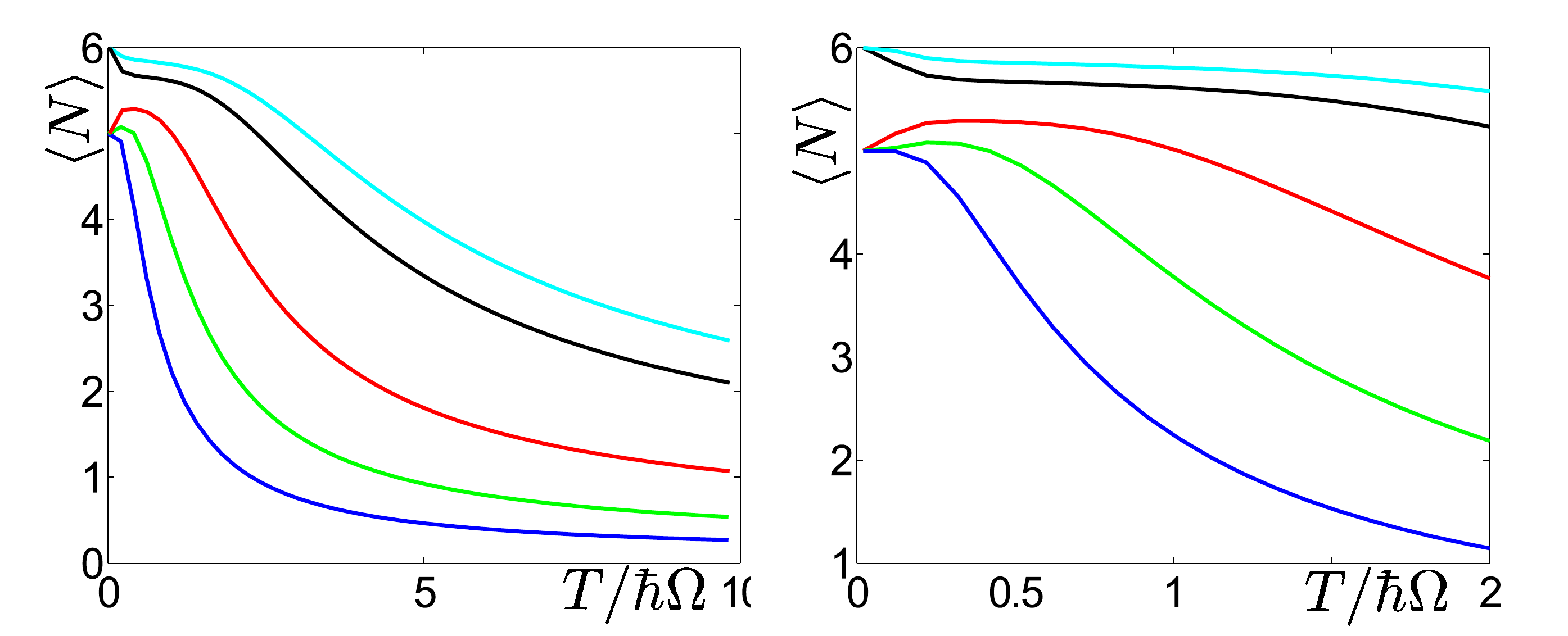}
\caption{Average dot population as a function of temperature. The left figure corresponds to parameters $\lambda^2=2$, $\alpha=0.01$, $E_C=1$ and from bottom to top $Q_g/e=0.05,\,0.1,\,0.2\,0.4$ and $0.5$. The right figure clarifies the low-temperature region.}
\label{ntemps}
\end{center}
\end{figure}

This physics is also clearly reflected in the average number $\langle N\rangle$ as function of temperature, as shown for various gate charges in Fig.\ \ref{ntemps}. For finite positive gate charge the zero temperature average value corresponds to the minimum of the right well. As the gate charge increases, the position of the minimum shifts and electrons are added to the dot in discrete jumps of one electron. At higher temperatures, the system is no longer trapped in the right well so that the average number of electrons decreases rapidly. However, the average number saturates at a finite value at high temperatures, reflecting the gate-induced asymmetry of the potential.

\begin{figure}[h]
\begin{center}
\includegraphics[width=0.99\columnwidth]{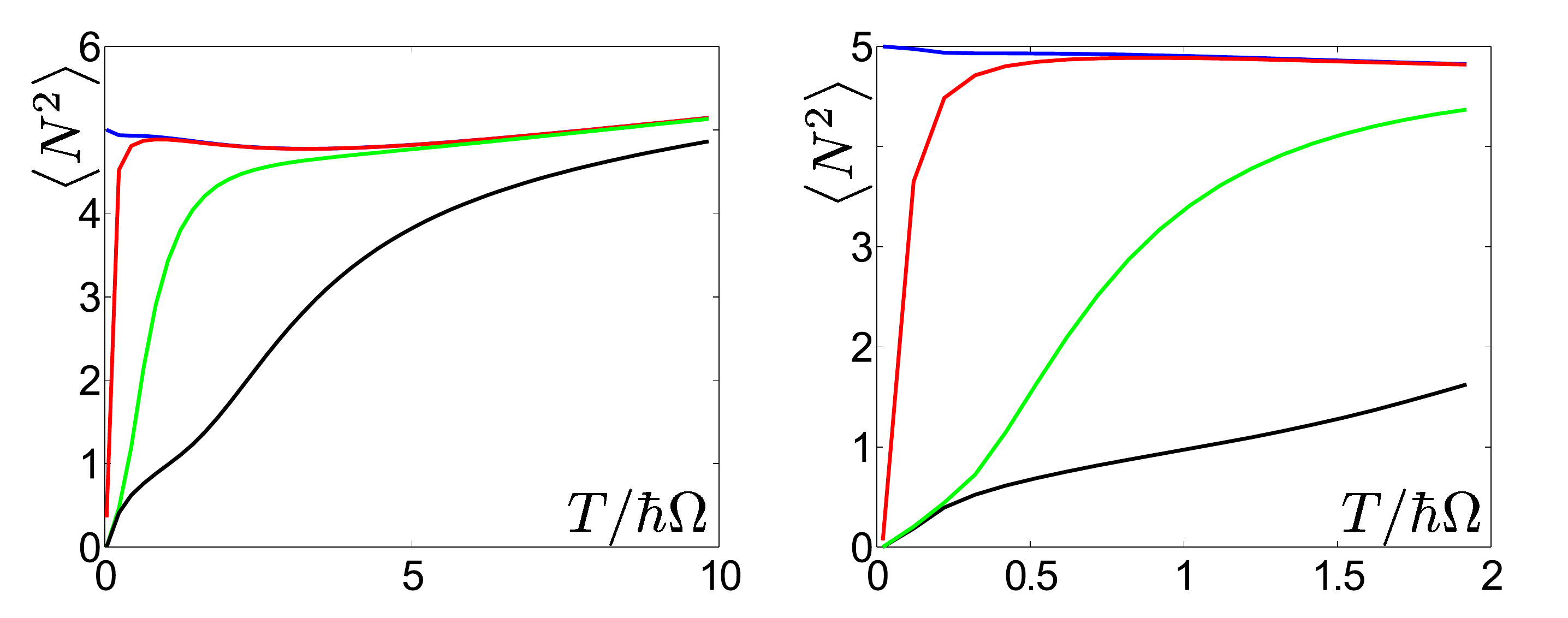}
\caption{Root mean square fluctuation as a function of temperature. The left figure corresponds to parameters $E_C=\hbar\Omega$, $\alpha=0.01(\hbar\Omega)^{-1}$, $\lambda^2=2$ and from bottom to top $Q_g/e=0.5,\,0.1,\,0.01,\, 0$. At vanishing gate the low-temperature fluctuations saturate at finite value, otherwise they will go to zero. The right figure clarifies the low-temperature region.}
\label{n2temps}
\end{center}
\end{figure}

Charge fluctuations also exhibit interesting behavior. In the absence of the electron-phonon interaction the electronic potential is $U(N)=E_c(N-Q_g/e)^2$, so the mean square number fluctuation
\begin{equation}\label{cfluc}
\Delta N^2= \langle N^2\rangle-\langle N\rangle^2
\end{equation}
is proportional to temperature, $\Delta N^2\propto T$. The situation is very different when the electrons couple to an anharmonic vibron mode. First of all, the zero-temperature number fluctuations do not vanish at zero gate charge, $Q_g=0$, but saturate to a finite value, see Fig.~\ref{n2temps}. This happens because there exist symmetric minima on both sides of the origin. The curve also exhibits a weak dip at finite temperatures since charge fluctuates more frequently towards values smaller than the charge at the minima. Eventually fluctuations grow due to the population of higher energy states. At finite gate values $Q_g\neq 0$ the fluctuations are drastically modified at low temperatures. The symmetry between the minima is broken and only states close to the preferred minimum are populated. When there is a single global minimum, the fluctuations vanish at zero temperature. At finite gate charge the curves approach to the zero gate curve at finite temperatures. At small gate charges, even a weak thermal excitation is sufficient to restore an approximate symmetry between the minima, leading to a sharp increase of the fluctuations from zero to the symmetric value at $Q_g=0$.

\section{Transport properties}
\label{transport}

\begin{figure}[h]
\begin{center}
\includegraphics[width=0.99\columnwidth]{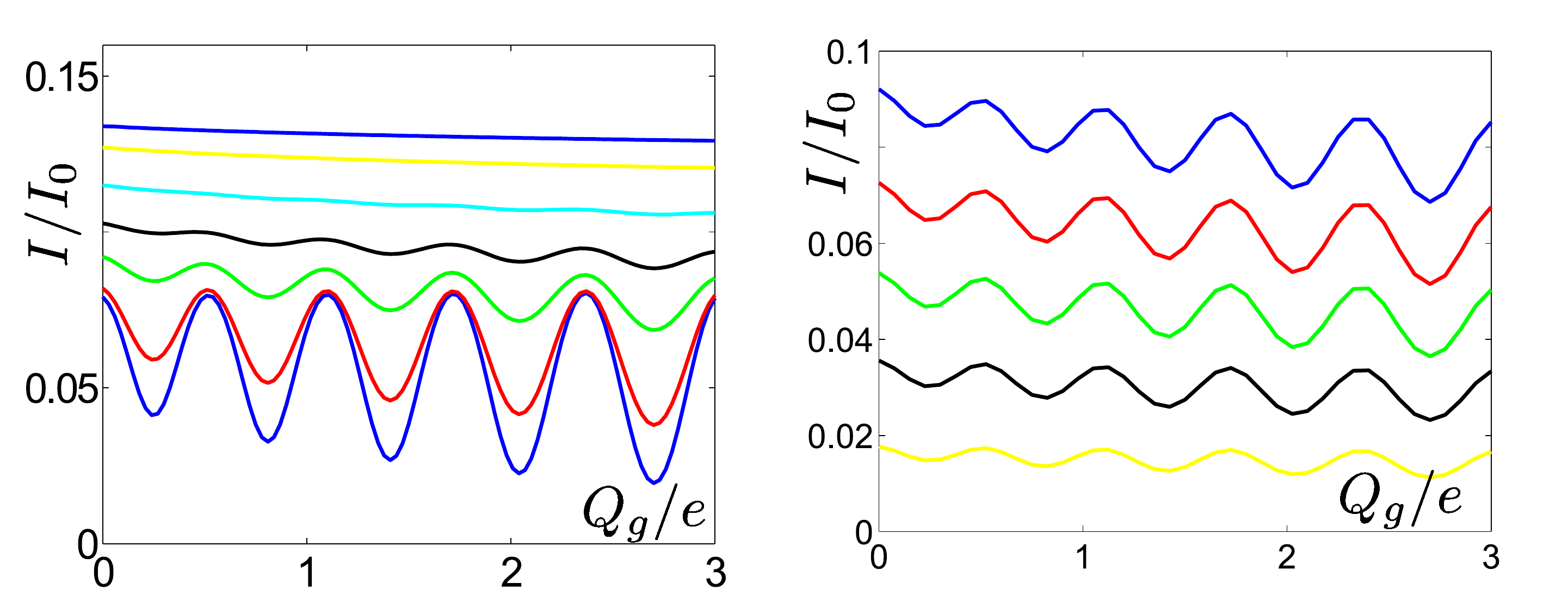}
\caption{Current as a function of the gate charge. Left: Different curves correspond to temperatures (from top to bottom) $T=0.2,\, 0.15,\, 0.1,\, 0.07,\, 0.05,\, 0.03,\, 0.02\times\hbar\Omega$, other parameters being $E_C=0.2\,\hbar\Omega$, $\alpha=0.05(\hbar\Omega)^{-1}$, $\lambda^2=0.5$, and $\mu_L=-\mu_R=0.1\hbar\Omega/e$. Right: Curves correspond to voltages (from top to bottom) $\mu_L=-\mu_R=0.1, 0.08, 0.06, 0.04, 0.02\times\hbar\Omega/e$, other parameters being $E_C=0.2\,\hbar\Omega$, $\lambda^2=0.5$, $\alpha=0.05$, and $T=0.05\,\hbar\Omega$. }
\label{cur}
\end{center}
\end{figure}

So far we have explored consequences of the negative-$U$ and the anharmonic phonon effects on equilibrium properties. In this section we focus on nonequilibrium characteristics of the system in the negative-$U$ regime. To simplify the analysis we assume a continuous density of states of the dot and that the energy scales separate $\hbar\Omega>>E_C$ so that only the lowest vibrational state $|E_0(N)\rangle$ is relevant at low temperatures and bias voltages. As implied by considerations in Section \ref{negative} and the estimate for the effective charging energy Eq.~(\ref{curvature}), close to the instability the curvature of the potential is weak and one expects the sequential tunneling processes to dominate the transport. In the lowest order of the tunneling coupling transport properties are described by a rate equation with Golden-Rule transition rates.\cite{beenakker} Rates for the processes in which the initial dot state contains $N$ electrons and where the electron number changes by one due to the tunneling through the left/right junction are
\begin{equation}\label{GR}
\Gamma_{N\pm1,N}^{\rm{L/R}}=\gamma^{\rm {L/R}}f\left(U^{\rm eff}(N\pm1)-U^{\rm eff}(N)\pm\mu_{\rm{L/R}}\right),
\end{equation}
where $f(x)=\frac{x}{e^{\beta x}-1}$, $\gamma^{\rm {L/R}}=\frac{|\langle E_0(N\pm1)|E_0(N)\rangle|^2}{e^2R_t}$ and $1/R_t=\frac{4\pi e^2|t|^2\nu_{\rm L/R}\nu_{\rm C}}{\hbar}$. In the positive-$U$ regime in the absence of the anharmonic term the matrix element $|\langle E_0(N\pm1)|E_0(N)\rangle|^2$ leads to the usual Franck-Condon suppression of the conductance given by $|\langle E_0(N\pm1)|E_0(N)\rangle|^2=e^{-\lambda^2}$. The probability $P(N)$ of having $N$ extra electrons on the dot follows from the detailed balance condition $\Gamma_{N-1,N}P(N)=\Gamma_{N,N-1}P(N-1)$, where the rates $\Gamma$ are given by the sum of the corresponding left and right lead rates. The stationary current through the SET can then be calculated from the expression
\begin{equation}\label{current}
I=-e\sum_N\left(\Gamma_{N+1,N}^{\rm{L}}-\Gamma_{N-1,N}^{\rm{L}} \right)P(N).
\end{equation}
 In Fig.~\ref{cur} we have plotted the current as a function of the gate charge at different temperatures and bias voltages. The current is normalized to the Franck-Condon suppressed tunneling current $I_0=e^{-\lambda^2}\hbar\Omega/eR_t $. At temperatures $T\gtrsim E_C/2$ current is a slowly varying function of the gate charge. Below this temperature it gradually starts to show signatures of Coulomb-like oscillations which become pronounced at low temperatures. Although this is in  qualitative agreement with the standard Coulomb blockade results in absence of the vibron interaction there are also some essential differences. All the curves exhibit a slowly decreasing tendency and the amplitude of current oscillations grow as a function of gate charge. The strict periodicity of the usual Coulomb-blockade conductance is broken by the nonlinear phonon interaction which does not leave the effective charging spectrum invariant as $Q_g$ is increased by multiples of the electron charge. Figure \ref{cur} (right) shows also that, within a reasonable accuracy, the current is a linear function of the applied voltage at low bias.

\begin{figure}[h]
\begin{center}
\includegraphics[width=0.99\columnwidth]{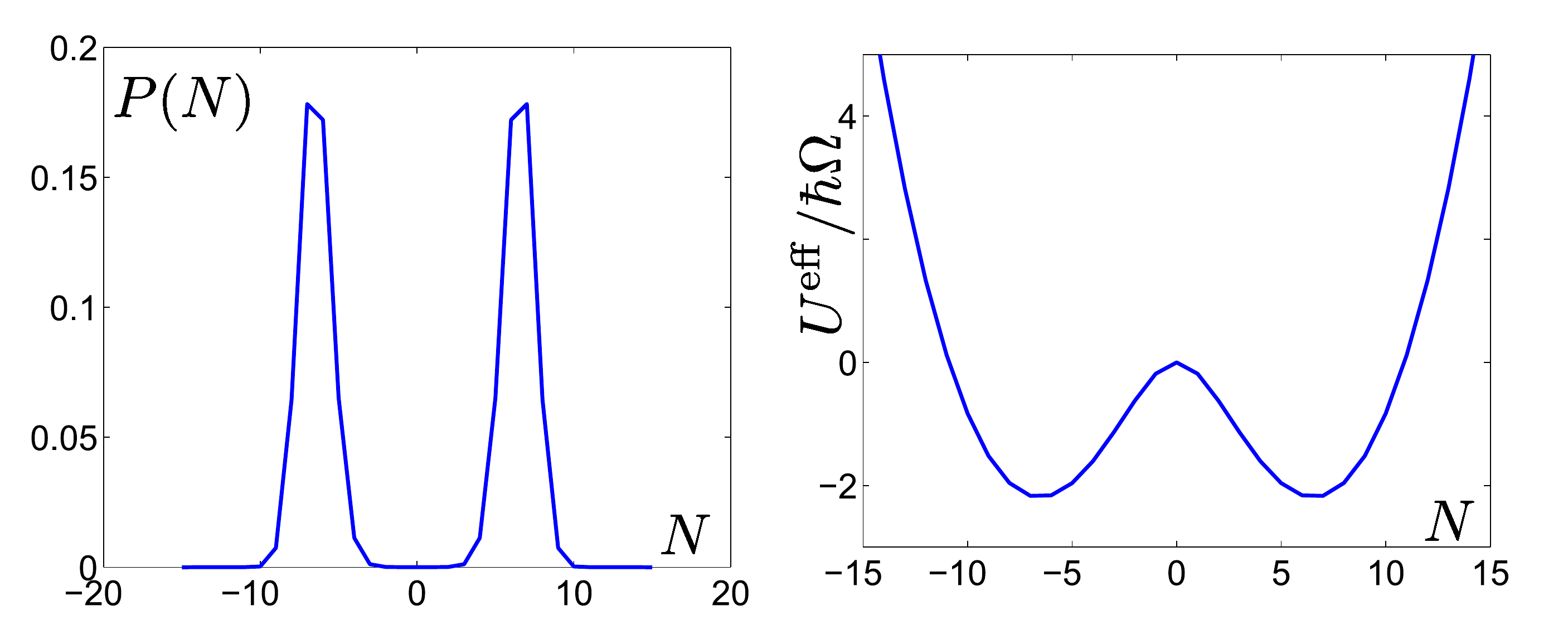}
\caption{Left: Probability distribution of the dot population corresponding to parameters $E_C=0.2\,\hbar\Omega$, $\alpha=0.05\,(\hbar\Omega)^{-1}$, $\lambda^2=0.5$ and $\mu_L=-\mu_R=0.1\,\hbar\Omega/e$, $Q_g=0$ at $T=0.1\,\hbar\Omega$. Right: The effective potential corresponding to the same parameters.}
\label{probab}
\end{center}
\end{figure}

It is noteworthy that the current is not completely blocked even at relatively low temperatures $T\sim 0.1\, E_C$ where the usual Coulomb-blockade current is completely suppressed around integer gate charges. The reduction of the blockade can be qualitatively understood by considering the estimate (\ref{curvature}), which indicates that in the vicinity of the instability the value of the effective charging is reduced. With parameters corresponding to Figs.~\ref{cur} and \ref{probab} the minima of the double-well potential are flatter than the minima corresponding to the bare electronic potential in the absence of vibrations and, as illustrated in Fig.~\ref{probab}, there exists two (and at weaker nonlinearity more) almost degenerate states in each well. Transitions between these closeby states at the bottom of the wells enable current to flow below temperatures where the bare electronic energy cost would block it. The estimate for the effective charging (\ref{curvature}) is valid only in the vicinity of the instability and by increasing $|E_C-\lambda\hbar\Omega|$ one eventually enters deep in the negative-$U$ region where the curvature at the potential minimum exceeds $E_C$. In this region the blockade is stronger than the usual Coulomb blockade and one has to consider higher-order tunneling processes.

\begin{figure}[h]
\begin{center}
\includegraphics[width=0.99\columnwidth]{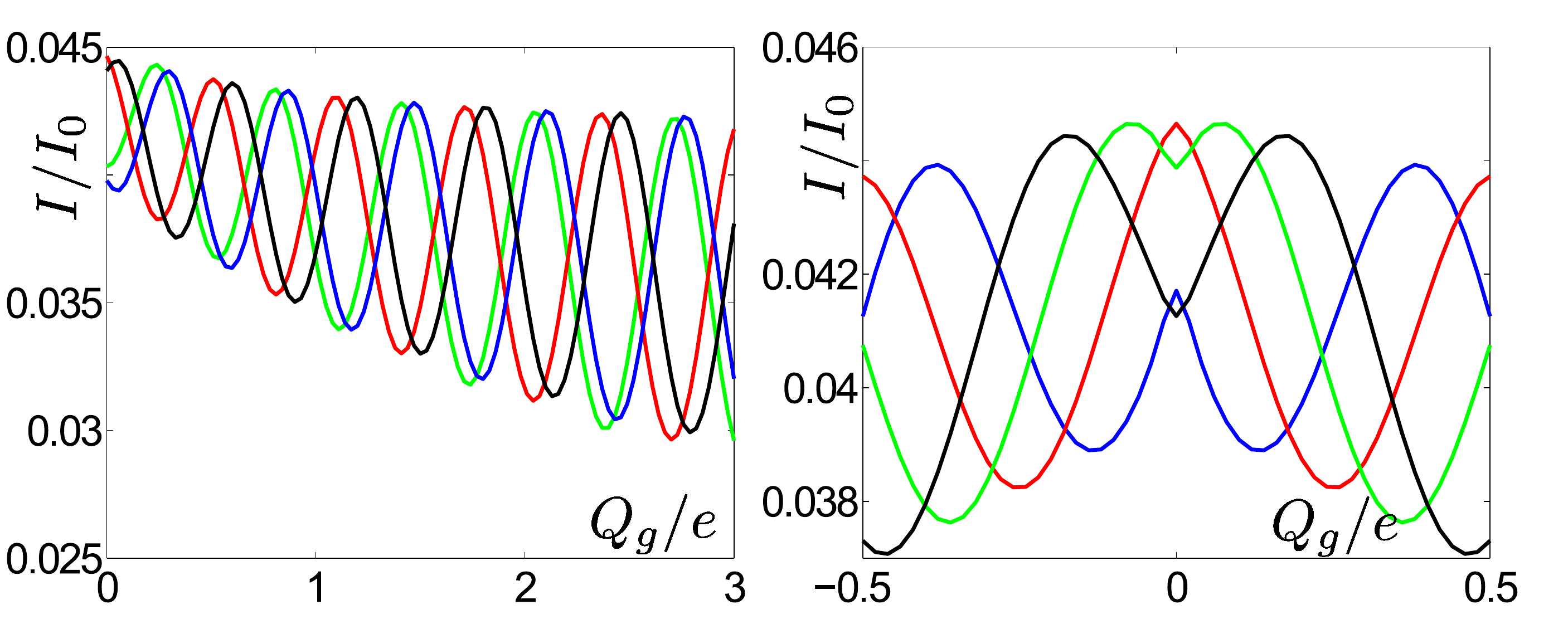}
\caption{Current as a function of the gate charge at different phonon nonlinearity strengths. Left: The different curves illustrate cases $\alpha=0.058\,(\hbar\Omega)^{-1}$ (green), $\alpha=0.05\,(\hbar\Omega)^{-1}$ (red), $\alpha=0.045\,(\hbar\Omega)^{-1}$ (blue) and $\alpha=0.04\,(\hbar\Omega)^{-1}$ (black),  other parameters being $E_C=0.2\,\hbar\Omega$, $\lambda^=0.5$, $T=0.05\,\hbar\Omega$ and $\mu_L=-\mu_R=0.05\,\hbar\Omega/e$. Right: Same quantities in the vicinity of $Q_g=0$. The curves correspond to $\alpha=0.056\,(\hbar\Omega)^{-1}$ (black), $\alpha=0.053\,(\hbar\Omega)^{-1}$ (green), $\alpha=0.05\,(\hbar\Omega)^{-1}$ (red) and $\alpha=0.047\,(\hbar\Omega)^{-1}$ (blue), other parameters as in the left plot.}
\label{curnon}
\end{center}
\end{figure}

In Fig.~\ref{curnon} we have plotted current vs. gate charge with a variable phonon nonlinearity. Small changes in the strength of the nonlinearity change energy differences of the nearby states at the potential minima leading to phase shifts in the current oscillations. Since the current is an even function of the gate charge it exhibits a cusp close to $Q_g=0$ depending on the phase of the current oscillations. The most pronounced effect is achieved when the current oscillations jump by half a period at the origin. At finite temperatures the cusp is always smooth and becoming sharper when temperature decreases. The reason for the existence of the cusp is the symmetry breaking between the two minima of the effective potential at finite $Q_g$. Even small gate values localize the system in one minimum leading to observable effects in the current through the structure.

\section{conclusions}
\label{conclusions}

In this paper, we studied effects of strong electron-phonon interaction in quantum dots with a vibrational degree of freedom. At sufficiently strong couplings the vibron-induced polaron shift overcomes the charging energy cost and the effective potential for electrons favors large charging. The instability towards an arbitrarily large electron population on the dot is regularized by the anharmonic contribution to the phonon energy. The effective potential differs qualitatively from the usual Coulomb repulsion, leading to characteristic modifications of the low temperature Coulomb blockade properties. Signatures include a rapid change in the average and the fluctuations of the electron number as function of the gate charge in the neighborhood of $Q_g=0$ as well as the temperature dependence of these quantities. Moreover, transport properties also show a number of unique signatures that can be used to characterize the negative-$U$ instability and the nonlinear phonon effects.

It is interesting to check whether in addition to molecules, an effectively negative $U$ could also be achieved in nanoelectromechanical systems. As shown by Eq.\ (\ref{cond2}), this requires an electron-vibron coupling (as measured by $\lambda\hbar\omega$) which is of the same order of magnitude as the charging and the single-particle energies. It turns out that this condition rules out currently existing suspended electron beam lithography samples. At the same time, the situation is more favorable (though still marginal) for suspended carbon-nanotube devices, for which strong electron-vibron coupling has been observed in several recent experiments.\cite{leroy,kuemmeth,huttel} The most likely vibron mode to cause a negative $U$ in this system is the radial breathing mode with frequency $\omega\sim c/L_\perp$. (Here, $c$ denotes the velocity of propagation of acoustic phonons in graphene and $L_\perp$ is the circumference of the nanotube.) The charging energy and the single-particle level spacing are of the same order (assuming a "fine-structure constant" $e^2/\hbar v_F\sim 1$ as appropriate for graphene) so that in order to reach the negative $U$ instability, the electron-vibron coupling $\lambda$ must exceed $[(v_F/c)(L_\perp/L)]^{1/2}$ for a nanotube of length $L$. Using $v_F/c\sim 10^2$, $L_\perp\sim 1$nm, and $L\sim 1\mu$m, we find that the critical coupling strength is $\lambda_c\sim 0.1-1$. Unfortunately, this coupling constant is relatively poorly understood at present. However, both experiment\cite{leroy} and theoretical estimates for one possible coupling mechanism\cite{mariani} suggest that it is of the same order as $\lambda_c$.\footnote{According to Ref.~\onlinecite{mariani}, $\lambda$ for the breathing mode scales as $L^{-\frac{1}{2}}$ which means that decreasing the critical coupling strength by increasing the length of the tube does not to improve the possibility of reaching the negative-$U$ regime. } While these estimates do not allow us to claim that the instability discussed in this paper could be observed in carbon-nanotube devices, they do suggest that realizing a negative $U$ in nanoelectromechanical systems is certainly conceivable.


\begin{acknowledgments}
We are grateful to E. Mariani, M. Raikh and Y. Gefen for valuable discussions. This work was supported in part by the DFG through Sfb 658 and SPP 1243 as well as through DIP. One of us (FvO) acknowledges the hospitality of the KITP at UCSB while part of this work has been performed. Work in Santa Barbara has been supported in part by the National Science Foundation under Grant No. PHY05-51164.
\end{acknowledgments}


\begin{thebibliography}{99}
\bibitem{note} Tunneling is still important in the sense that in the equilibration process the island can change its charging state by exchanging electrons with the leads.
\end{thebibliography}
\end{document}